\documentclass[letter,bibyear]{aa} 

\usepackage{lscape}
\usepackage{float}
\usepackage{graphicx}
\usepackage{txfonts}
\begin{document} 

\title{On the Kennicutt-Schmidt scaling law of submillimetre galaxies}

   \author{O.~Miettinen\inst{1}, I.~Delvecchio\inst{1}, V.~Smol\v{c}i\'{c}\inst{1}, M.~Aravena\inst{2}, D.~Brisbin\inst{2}, and A.~Karim\inst{3}}

   \institute{Department of Physics, Faculty of Science, University of Zagreb, Bijeni\v{c}ka cesta 32, HR-10000 Zagreb, Croatia \\ \email{oskari@phy.hr} \and 
N\'ucleo de Astronom\'{\i}a, Facultad de Ingenier\'{\i}a, Universidad Diego Portales, Av. Ej\'ercito 441, Santiago, Chile \and 
Argelander-Institut f\"{u}r Astronomie, Universit\"{a}t Bonn, Auf dem H\"{u}gel 71, D-53121 Bonn, Germany}

   \date{Received ; accepted}

\authorrunning{Miettinen et al.}
\titlerunning{The K-S relation of SMGs}

\abstract{The star formation rate (SFR) per unit area correlates well with the gas surface density for different types 
of galaxies. However, this Kennicutt-Schmidt (K-S) law has not yet been examined for a large, homogeneously selected sample 
of submillimetre galaxies (SMGs), which could provide useful SF implementation information for models of massive galaxy formation 
and evolution.}{We aim at determining the K-S law parameters for the first time for a well-selected, statistical sample of SMGs.}{We used 
the Atacama Large Millimetre/submillimetre Array (ALMA) to conduct a high resolution ($0\farcs2$), 870~$\mu$m continuum imaging 
survey of 40 SMGs, which were initially selected at 1.1~mm in the COSMOS field. We analysed a sample of 32 out of the 40 target SMGs, 
for which our new ALMA 870~$\mu$m data provide information about the spatial extent of dust emission, and all of which have dust-obscured SFR and 
dust-based gas mass estimates available from our previous study.}{We divided our sample into equally large subsamples of main-sequence (MS) objects and starbursts 
(factor of $>3$ above the MS), and found their K-S relations to be of the form $\Sigma_{\rm SFR} \propto \Sigma_{\rm gas}^{0.81\pm0.01}$ and 
$\Sigma_{\rm SFR} \propto \Sigma_{\rm gas}^{0.84\pm0.39}$, respectively.}{The slightly sub-linear K-S slopes we derived suggest that the SF efficiency (SFE) 
is nearly constant across the $\Sigma_{\rm gas}$ range probed. Under the assumption of a Galactic CO-to-H$_2$ conversion factor ($\alpha_{\rm CO}$) for 
the whole sample, the MS SMGs obey a constant global SFE of about $21\%$ per 100~Myr, while that of starburst SMGs is about $27\%$ per 100~Myr. 
The corresponding gas depletion times are $\sim480$~Myr and $370$~Myr. On average, our SMGs have $\Sigma_{\rm gas}\gtrsim10^{3.9}$~M$_{\sun}$~pc$^{-2}$, 
which suggests that they are Eddington-limited. This is consistent with the theoretical expectation of a linear K-S relation for such systems. 
However, size measurements of the CO-emitting regions of SMGs, and the $\alpha_{\rm CO}$ values of SMGs are needed to further constrain their $\Sigma_{\rm gas}$ values.}

\keywords{Galaxies: evolution -- Galaxies: formation -- Galaxies: starburst -- Galaxies: star formation -- Submillimetre: galaxies}

   \maketitle
%

\section{Introduction}

The empirical Kennicutt–Schmidt (K-S) law quantifies the amount of cold interstellar gas 
required to sustain a given star formation rate (SFR; \cite{schmidt1959}; \cite{kennicutt1998}, 
hereafter K98; see \cite{kennicutt2012} for a review). Specifically, K98 found that 
the galaxy-integrated SFR surface density ($\Sigma_{\rm SFR}$) and total (atomic plus molecular, 
$\ion{H}{I}+{\rm H}_2$) gas surface density ($\Sigma_{\rm gas}$) of normal star-forming disk galaxies 
and luminous infrared (IR) selected starbursts are tightly linked to each other over about five decades 
in $\Sigma_{\rm gas}$ through a functional form of $\Sigma_{\rm SFR}\propto \Sigma_{\rm gas}^{1.4\pm0.15}$. 
Besides normal spirals and starbursts studied by K98, the K-S-type star-formation relations have been explored 
for different types of galaxies with different physical properties, 
such as low-surface-brightness galaxies (\cite{wyder2009}) and luminous IR galaxies (\cite{garcia2012}). 
However, studies of the K-S law of the most intensely star-forming dusty galaxies, 
the so-called submillimetre galaxies (SMGs; see \cite{casey2014} for a review), 
are not only few in number, but they have also been based on small, heterogenous, 
and partly overlapping literature samples (\cite{bouche2007}; \cite{daddi2010b}, hereafter D10b; 
\cite{bothwell2010}; \cite{genzel2010}; see also \cite{hodge2015}). For instance, D10b found that 
while SMGs and normal disks have a common K-S slope of 1.42, 
which is fully consistent with the K98 value, the SMGs occupy a higher $\Sigma_{\rm SFR}$ regime 
of the K-S diagram with 0.9~dex higher normalisation.  
This is considered an indication that SMGs, which are potentially driven by gas-rich mergers, 
are relatively more efficient star formers (see also \cite{genzel2010}, 2015). 

Inherently, the observed galactic scale K-S relation is a manifestation of the low global SF efficiency (SFE). 
Although the exact parameters of the K-S relation are dependent on several factors (e.g. the SFR and gas tracers used; e.g. 
\cite{krumholz2007}; \cite{liu2011}; \cite{momose2013}), the global SFE appears to be only a few percent 
(e.g. K98; \cite{bigiel2008}; \cite{genzel2010}). In this regard, to better understand the overall role played by SMGs 
in the formation and evolution of massive galaxies, it is pivotal to try to quantify 
how efficiently SMGs turn their gas into stars, yet this requires an analysis of a well-selected statistical source sample.  

In this Letter, we report our results regarding the K-S law of SMGs, which were detected at 870~$\mu$m with the Atacama Large Millimetre/submillimetre Array (ALMA). This represents the first homogenous, statistically more significant sample of SMGs for which the K-S law has been explored so far. The SMG sample and observations are described in Sect.~2, while the analysis and results are described and discussed in Sect.~3. Section~4 summarises our results. Throughout this Letter, we adopt a Chabrier (2003) initial mass function (IMF), and assume a $\Lambda$CDM (Lambda cold dark matter) cosmology with the dark energy density $\Omega_{\Lambda}=0.70$, and total matter density $\Omega_{\rm m}=0.30$, while the Hubble constant is set at $H_0=70$~km~s$^{-1}$~Mpc$^{-1}$. 

\section{Source sample and ALMA observations}

The target SMGs, called AzTEC/C1--C27, were originally uncovered by the AzTEC $\lambda_{\rm obs}=1.1$~mm blank-field continuum survey of the inner 
0.72~deg$^2$ of the COSMOS field (\cite{aretxaga2011}). The sources AzTEC/C1--C27 correspond to a signal-to-noise limited subsample of the AzTEC 
single-dish sources with ${\rm S/N}_{\rm 1.1\, mm}^{\rm AzTEC}\geq5.5$ ($S_{\rm 1.1\, mm}=5.7-13$~mJy), and were observed as part of our ALMA follow-up 
survey in Cycle~2 at $\lambda_{\rm obs}=1.3$~mm and $\sim1\farcs6 \times 0\farcs9$ resolution (PI: M.~Aravena; Aravena et al., in prep.). 
The dedicated ALMA pointings towards these 27 AzTEC sources revealed 41 sources altogether, 
at a S/N$_{\rm 1.3\, mm}^{\rm ALMA}\geq5$ ($S_{\rm 1.3\, mm}=0.55-7.25$~mJy).

We followed up the 1.3~mm sources detected towards AzTEC/C1--C27 with ALMA in Cycle~4 using Band~7 continuum observations at $\lambda_{\rm obs}=870$~$\mu$m under project {\tt 2016.1.00478.S} (PI: O.~Miettinen). The observations were carried out on 28 October 2016. Altogether, 40 ALMA 1.3~mm sources were covered by 34 pointings ($16\farcs7$ FWHM field-of-view), with a total on-source integration time of about 1.3~min per pointing (AzTEC/C3b was not observed). The observations were made using the 12~m array with 41 antennas, where the baselines ranged from 18.6~m (21.3~k$\lambda$) to 1.1~km (1\,260~k$\lambda$). The large number of antennas allowed us to reach an excellent $uv$-coverage even in the aforementioned short integration time. The amount of precipitable water vapour was only about 0.38~mm. The phases were calibrated by observations of the Seyfert 1 galaxy J0948+0022, while the BL Lac object J1058+0133 was observed for amplitude and bandpass calibration. The correlator was configured in four spectral windows centred at 336.5~GHz and 338.5~GHz in the lower sideband, and at 348.5~GHz and 350.5~GHz in the upper sideband, each covering a bandwidth of 1.875~GHz divided into 128 channels of 15.625~MHz (with dual polarisation). Hence, the total bandwidth available for continuum observations was 7.5~GHz.

The visibility data were edited, calibrated, and imaged using the standard ALMA pipeline of the Common Astronomy Software Applications (CASA; \cite{mcmullin2007}) version 4.7.0. The final images were created using the {\tt tclean} task by adopting Briggs weighting with a {\tt robust} parameter of 0.5. The resulting images have a typical (median) synthesised beam of $0\farcs192 \times 0\farcs176$, while the typical $1\sigma$ rms noise of the final images is 0.155~mJy~beam$^{-1}$, which was estimated from emission-free regions after correction for the primary beam (PB) response. 

Out of the 40 target sources, 36 were detected with a S/N ratio ranging from 5.9 to 33 (see Fig.~\ref{figure:maps}). 
The four sources that were not detected are AzTEC/C1b, C8b, C10c, and C13b (S/N$_{\rm 1.3\, mm}^{\rm ALMA}=5.2$, 5.5, 5.1, and 10.2, respectively). 
A potential reason for these non-detections is that the emission was resolved out at $0\farcs2$ resolution. 
To test this possibility, we convolved the images with a Gaussian smoothing kernel of different radii. 
No emission was recovered towards AzTEC/C1b ($\sim5\farcs6$ south-west (SW) of the phase centre (PC)) and C10c 
(source at the PC), which suggests that these sources might be spurious. Indeed, AzTEC/C1b and C10c have 
no multiwavelength counterparts, unlike C8b and C13b (\cite{brisbin2017}). Also, the map smoothing did not reveal 
any clear source at the 1.3~mm position of AzTEC/C13b, and in this case the non-detection might be caused by PB attenuation, 
because the source lies $\sim6\farcs4$ to the SW of the PC, where the map starts to become noisy. 
However, although AzTEC/C8b also lies near the noisy map edge ($\sim7\arcsec$ to the SW from the PC), 
the source appeared in smoothed images (starting to become visible at $0\farcs30 \times 0\farcs25$ resolution, 
where the corresponding map rms noise is $\sim0.2$~mJy~beam$^{-1}$) with a hint of two components of 
$5\sigma$ and $4.7\sigma$ significance separated by $0\farcs26$. AzTEC/C8b also has a large radio-emitting full width at half maximum (FWHM) 
size of $1\farcs7 \times 1\farcs1$ (\cite{miettinen2017a}, hereafter M17a), which is consistent with the finding that its dust-emitting region 
was resolved out. Owing to the location of C8b near the noisy map boundary, and the fact that it was resolved out at $0\farcs2$ resolution,  
we do not consider it in the subsequent analysis to preserve the homogeneity of the data set.

\section{Data analysis, results, and discussion}

An integral part of the present analysis is to determine the spatial scale of the observed-frame 870~$\mu$m 
emission. For this purpose, we used the NRAO Astronomical Image Processing System (AIPS) software package. 
Specifically, the beam-deconvolved (intrinsic) sizes were derived through two-dimensional 
elliptical Gaussian fits to the image plane data using the AIPS task {\tt JMFIT}. 
The Gaussian fitting was performed inside a rectangular box enclosing the source, and the fit was 
restricted to the pixel values of $\geq 2.5\sigma$. 

In the subsequent analysis, we use the deconvolved major axis FWHM as the diameter of the source, because the major axis 
represents the physical source extent in the case of isotropically oriented disks. 
All the sources were resolved along the major axis; the deconvolved FWHM was always 
found to be larger than one-half the synthesised beam major axis FWHM (see Table~\ref{table:sample}). 
The median value of ${\rm FWHM_{\rm maj}}$ is $0\farcs31^{+0.15}_{-0.10}$ ($2.4^{+1.1}_{-0.8}$~kpc), where 
the uncertainty represents the 16th--84th percentile range. This is in good agreement with previous 
studies of SMG sizes measured through ALMA 870~$\mu$m observations (\cite{simpson2015}; \cite{hodge2016}), although 
the source is not always well modelled with an elliptical Gaussian profile (Fig.~\ref{figure:maps}). 
As a consistency check, we also used CASA to determine the source sizes (the {\tt imfit} task), 
and found very good agreement with our AIPS/{\tt JMFIT} results, the mean (median) ratio between 
the two being $\langle {\rm Size(AIPS)}/{\rm Size(CASA)}\rangle=1.06$ (1.02).

The source radius, which enters into the calculation of the surface densities, 
was defined as $R=0.5\times {\rm FWHM_{\rm maj}}$, which is appropriate for a circular disk. 
Both the SFR and gas mass ($M_{\rm gas}$) values were adopted from Miettinen et al. (2017b, hereafter M17b), 
who used the latest version of {\tt MAGPHYS} (\cite{dacunha2015}) to fit the panchromatic spectral energy distributions (SEDs) 
of the target SMGs. The number of SMGs that have both the SED and size information available is 32, 
and their redshifts range from $z=1.1^{+2.6}_{-1.1}$ to $z=5.3^{+0.7}_{-1.1}$ (40.6\% are spectroscopically confirmed, 
while the remaining redshifts are photometric; \cite{brisbin2017}). 

The best-fit {\tt MAGPHYS} SEDs were integrated over the rest-frame wavelength range of 
$\lambda_{\rm rest}=8-1\,000$~$\mu$m to derive the IR luminosities ($L_{\rm IR}$). The values of $L_{\rm IR}$ were then used to estimate 
the dust-obscured, 100~Myr averaged SFR using the K98 relationship. 

The gas masses were estimated using the Scoville et al. (2016) 
calibration and employing the ALMA 1.3~mm flux densities of the sources. These dust-based $M_{\rm gas}$ values refer to the molecular (H$_2$) 
gas mass (see M17b for further details). We note that similar to the canonical K-S relation (K98), 
which assumes a Galactic CO-to-H$_2$ conversion factor ($\alpha_{\rm CO}$) for both the normal disks and starbursts, 
the Scoville et al. (2016) method is calibrated using a comparable, single Galactic $\alpha_{\rm CO}$ of 
6.5~M$_{\sun}$~(K~km~s$^{-1}$~pc$^2$)$^{-1}$ (including the helium contribution) for different types of star-forming galaxies, including SMGs. 

 We also note that only two of our target sources, AzTEC/C5 and C17, have CO-inferred $M_{\rm gas}$ estimates available, and when the different 
assumptions about $\alpha_{\rm CO}$ are taken into account, they agree within a factor of two with our dust-based values (being either lower or higher; 
we refer to M17b, and references therein). 

Finally, because the source sizes we derived refer to the FWHM extent, the surface densities were calculated as $\Sigma_{\rm SFR}={\rm SFR}/(2\pi R^2)$ and 
$\Sigma_{\rm gas}=M_{\rm gas}/(2\pi R^2)$. The associated uncertainties were propagated from the uncertainties in SFR, $M_{\rm gas}$, and size.

\begin{figure*}
\begin{center}
\includegraphics[scale=0.46]{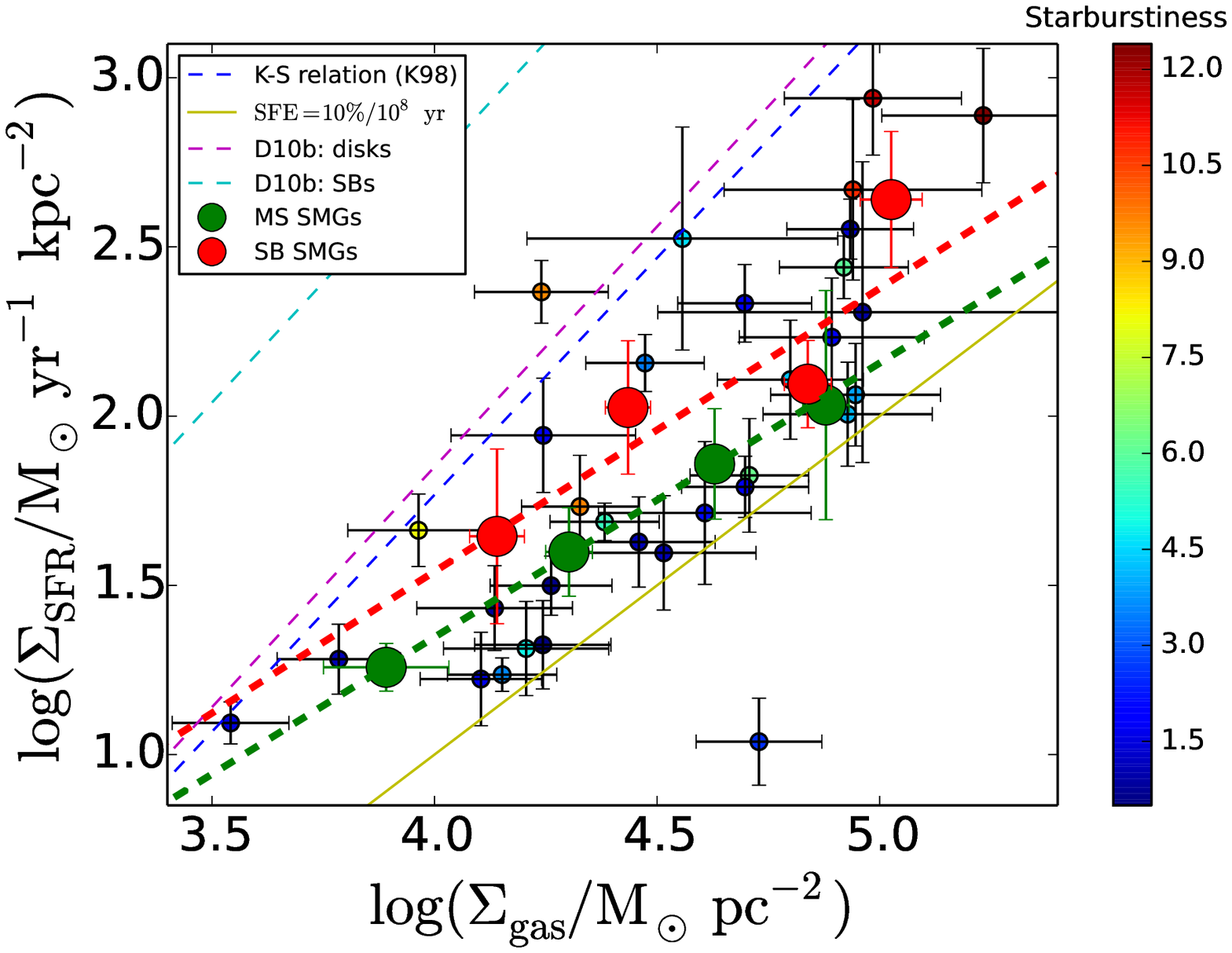}
\includegraphics[scale=0.43]{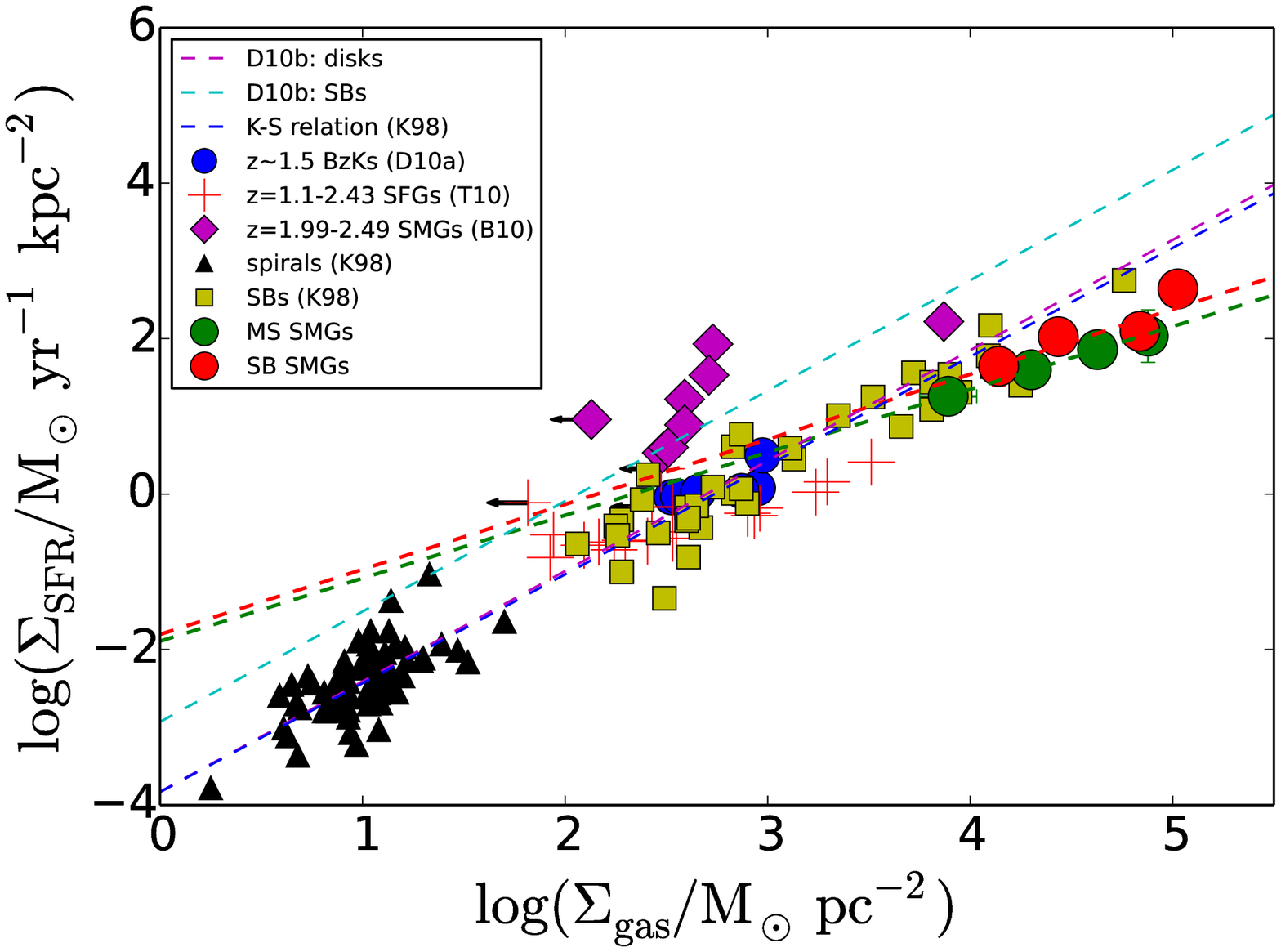}
\caption{\textbf{Left:} Kennicutt-Schmidt diagram for the target SMGs. The individual data points are colour-coded with the distance from 
the Speagle et al. (2014) MS as shown in the colour-bar on the right. The green and red filled circles represent the mean values of the binned
MS and starburst data, where the latter population is defined as lying above the MS by a factor of $>3$. Each bin contains four SMGs, 
and the error bars represent the standard errors of the mean values (see Table~\ref{table:data}). The green and red dashed lines represent the least squares fits to the binned data sets, the blue dashed line shows the K98 relationship, and the magenta and cyan dashed lines show the D10b relations for disks and starbursts, respectively. For reference, the yellow solid line corresponds to a constant global SFE of $10\%$ per 100~Myr, which corresponds to a gas depletion time of $\tau_{\rm dep}=1$~Gyr. \textbf{Right:} The binned averages from the left panel compared with selected literature studies. The black triangles and yellow squares show the spiral galaxy and starburst data from K98, respectively, the red plus signs show the $z=1.10-2.43$ star-forming galaxy data from Tacconi et al. (2010), the blue circles show the $z=1.414-1.6$ BzK-selected disk galaxies from Daddi et al. (2010a), and the magenta diamonds represent the $z=1.21-2.49$ SMG data from Bothwell et al. (2010). The arrows pointing left indicate upper limits to $\Sigma_{\rm gas}$. The dashed lines have the same meaning as in the left panel.}
\label{figure:ks}
\end{center}
\end{figure*}

The K-S diagram of our SMGs is shown in the left panel in Fig.~\ref{figure:ks}, while our data are compared with literature studies in the right panel of the figure. The individual sources are colour-coded according to the distance from the main sequence (MS) as defined by Speagle et al. (2014). We also show the binned version of the data, where the sample was divided into MS objects and super-MS objects or starbursts (defined to be offset from the MS mid-line by a factor of $>3$; see M17b). The linear least squares fits ($\log \Sigma_{\rm SFR}= a\times \log \Sigma_{\rm gas}+b$)  through the binned data points yielded the slope and $y$-intercept of $(a=0.81\pm0.01,\,b=-1.89\pm0.05)$ for the MS SMGs, and $(a=0.84\pm0.39,\,b=-1.81\pm1.84)$ for the starburst SMGs. The quoted uncertainties in the fit parameters represent the $1\sigma$ standard deviation errors, and they were derived from the $\Sigma_{\rm SFR}$ uncertainties. As illustrated in Fig.~\ref{figure:ks}, our SMG $\Sigma_{\rm SFR}-\Sigma_{\rm gas}$ relations have flatter slopes and higher zero points than the K98 relation and the D10 relationships for normal disks and starbursts (with different $\alpha_{\rm CO}$ values), where the former is very similar to the canonical K98 relation. However, our SMGs have extreme gas surface densities of $\Sigma_{\rm gas}\gtrsim10^{3.9}$~M$_{\sun}$~pc$^{-2}$ on average, and hence we are mostly probing a different $\Sigma_{\rm gas}$ regime than K98 and D10 (but using dust rather than CO to estimate $M_{\rm gas}$). Such high densities make the gaseous interstellar medium (ISM) highly optically thick even in the re-radiated IR, and the radiation pressure on dust grains makes the system become Eddington-limited (e.g. \cite{ballantyne2013}; \cite{thompson2016}, and references therein). Interestingly, the K-S slope for the radiation-pressure-supported, Eddington-limited disk is expected to be unity (the stellar radiative flux $F_{\star}\propto \Sigma_{\rm SFR}$, and the Eddington flux $F_{\rm Edd}\propto \Sigma_{\rm gas}$; \cite{thompson2005}; \cite{ostriker2011}), which is broadly consistent with our results, particularly for starburst SMGs for which the K-S slope is consistent with unity within $\sim0.4\sigma$.

Our best-fit scaling relations shown in Fig.~\ref{figure:ks} suggest that the SFE is fairly weakly dependent on $\Sigma_{\rm gas}$ at the high 
densities probed (${\rm SFE}=\Sigma_{\rm SFR}/\Sigma_{\rm gas}\propto \Sigma_{\rm gas}^{-0.19\pm0.01}$ for the MS SMGs, and ${\rm SFE}\propto \Sigma_{\rm gas}^{-0.16\pm0.39}$ above the MS). To estimate the global SFEs of our SMGs, we fit the binned data with slopes constrained to unity. On average, our MS SMGs are consistent with a constant global SFE of $21^{+2}_{-1}\%$ per 100~Myr, while that for our starburst SMGs is $27^{+6}_{-6}\%$ per 100~Myr. The corresponding gas depletion times are $\tau_{\rm dep}={\rm SFE}^{-1}\simeq 480^{+20}_{-45}$~Myr and $\simeq370^{+106}_{-67}$~Myr, respectively. 

If the gas scale heights ($h\propto \Sigma_{\rm gas}/\rho_{\rm gas}$, where $\rho_{\rm gas}$ is the gas volume density) do not change much among different sources, the K-S law $\Sigma_{\rm SFR}\propto \Sigma_{\rm gas}^{1.4\pm0.15}$ (K98) is consistent with $\rho_{\rm SFR}\propto \rho_{\rm gas}/\tau_{\rm ff}\propto\rho_{\rm gas}^{1.5}$, where $\tau_{\rm ff}\propto \rho_{\rm gas}^{-0.5}$ is the free-fall timescale. Hence, a possible interpretation is that the K-S relation is a manifestation of star formation being predominantly driven by large-scale gravitational disk instabilities with a characteristic dynamical (fragmentation) timescale given by that of free-fall collapse (e.g. \cite{kennicutt1989}; \cite{elmegreen2002}). The K-S relations and $\tau_{\rm dep}(\Sigma_{\rm gas})$ dependencies we derived are shallower than what would be expected from this free-fall paradigm, which could reflect the fact that our measurements are averaged over entire SMGs, and are hence expected to be sensitive to fairly similar ISM characteristics across the sample (e.g. \cite{krumholz2007}; \cite{bigiel2008}). 

There are a number of critical assumptions (e.g. $\alpha_{\rm CO}$) and caveats in the above analysis. For example, a lower value of $\alpha_{\rm CO}=0.8$~M$_{\sun}$~(K~km~s$^{-1}$~pc$^2$)$^{-1}$, which is often adopted for ultraluminous infrared galaxies (ULIRGs; \cite{downes1998}), might be more appropriate for SMGs than a Galactic value. In Fig.~\ref{figure:ksulirg}, we show two alternative K-S diagrams, one derived by assuming the aforementioned ULIRG $\alpha_{\rm CO}$ factor for all of our sources, and another one with a bimodal $\alpha_{\rm CO}$ distribution, namely a ULIRG-like value for the starburst SMGs, and the same Galactic value for the MS objects as in Fig.~\ref{figure:ks}. We stress that these different assumptions about the $\alpha_{\rm CO}$ value do not influence the K-S slope values quoted above, only the normalisations (see Appendix~B). 

Another caveat is that the dust-emitting sizes of SMGs are found to be more compact that the spatial extent of their molecular gas reservoir (see M17a, and references therein), and hence our $\Sigma_{\rm gas}$ values could well be overestimated. On the other hand, M17a found that the observed-frame 3~GHz radio-emitting sizes of the target SMGs (see Fig.~\ref{figure:maps}) have a median value comparable to that of the CO-emitting gas component measured through mid-$J$ rotational transitions by Tacconi et al. (2006) for their sample of SMGs (consistent with the SMGs' 1.4~GHz and CO sizes studied by Bothwell et al. (2010)). Hence, one might think that the extent of radio emission is a better estimate of the distribution of molecular gas than the rest-frame far-IR emission. However, it should be noted that the molecular gas reservoir of SMGs probed through CO$(J=1-0)$ observations is found to be more extended than the denser and warmer component giving rise to mid-$J$ CO emission (we refer to the discussion in M17a). Nevertheless, we also derived the $\Sigma_{\rm gas}$ values using the radio sizes from M17a, and constructed another version of the K-S diagram, which is shown in the top panel in Fig.~\ref{figure:ks_radio} (the bottom panel has also $\Sigma_{\rm SFR}$ calculated over the 3~GHz size). In this case, we derived a highly sublinear ($a=0.40\pm0.07$) and even negative slope $(a=-0.16\pm0.02$) for the MS and super-MS objects, which suggests that the 3~GHz sizes are not universally representative of our SMGs' molecular gas extent. This raises the question of which size scale is the most appropriate to compute $\Sigma_{\rm gas}$, and if the CO emission size is used, then which CO transition is the most relevant: $J=1-0$ to probe the full, diffuse molecular gas component, or a higher $J$ transition, which arises from a denser and warmer gas associated with an on-going star formation. Our results are in line with K98, who suggested that it is vital to correlate the values of $\Sigma_{\rm SFR}$ and $\Sigma_{\rm gas}$ over regions co-equal in size. 

\section{Summary and conclusions}

We used ALMA to carry out a $0\farcs2$ resolution, 870~$\mu$m continuum imaging survey of a sample of SMGs in COSMOS. When combined with 
the source size information provided by these observations, our previous dust-based SFR and gas mass estimates for these sources allowed us to examine their K-S type, $\Sigma_{\rm SFR}-\Sigma_{\rm gas}$ scaling law. The dust-inferred $M_{\rm gas}$ values used in the analysis are based on the critical assumption of a uniform Galactic CO-to-H$_2$ conversion factor. We found that the average relationships for our MS and starburst SMGs are $\Sigma_{\rm SFR}\propto\Sigma_{\rm gas}^{0.81\pm0.01}$ and $\Sigma_{\rm SFR}\propto\Sigma_{\rm gas}^{0.84\pm0.39}$. The MS SMGs are consistent with an average constant global SFE of about $21\%$ per 100~Myr, while that of starburst SMGs is somewhat higher, about $27\%$ per 100~Myr. These SFEs correspond to gas consumption times of $\sim480$~Myr and $370$~Myr, respectively. The gas surface densities of the studied SMGs are typically $\Sigma_{\rm gas}\gtrsim10^{3.9}$~M$_{\sun}$~pc$^{-2}$, which suggest that the sources exceed the Eddington limit from radiation pressure on dust. Moreover, the slightly sub-linear, or quasi-linear $\Sigma_{\rm SFR}-\Sigma_{\rm gas}$ relations we derived are in broad agreement with the theoretical expectation of the SFR and gas surface densities being linearly correlated with each other for the radiation pressure supported, Eddington-limited disk. Our study also demonstrates how the source size can be one of the major bottlenecks in deriving the K-S law of SMGs, and this warrants further observations of the gas distribution in these galaxies.

\begin{acknowledgements}

We thank our referee for constructive comments and useful suggestions that 
helped us improve this paper. We also thank Dirk Petry for his help with our ALMA Cycle~4 data products. 
This research was funded by the European Union's Seventh Framework programme 
under grant agreement 337595 (ERC Starting Grant, 'CoSMass'). M.~A. acknowledges 
partial support from FONDE-CYT through grant 1140099. A.~K. acknowledges support 
by the Collaborative Research Council 956, sub-project A1, funded by 
the Deutsche Forschungsgemeinschaft (DFG). This paper makes use of the following ALMA data: ADS/JAO.ALMA\#2016.1.00478.S. 
ALMA is a partnership of ESO (representing its member states), 
NSF (USA) and NINS (Japan), together with NRC (Canada), NSC and ASIAA (Taiwan), 
and KASI (Republic of Korea), in cooperation with the Republic of Chile. 
The Joint ALMA Observatory is operated by ESO, AUI/NRAO and NAOJ. 

\end{acknowledgements}

\appendix

\section{ALMA 870~$\mu$m images, the dust-emitting sizes, and the average gas and SFR surface densities}

\begin{table}[H]
\renewcommand{\footnoterule}{}
\caption{Source sample and the sizes derived through Gaussian fits.}
{\scriptsize
\begin{minipage}{1\columnwidth}
\centering
\label{table:sample}
\begin{tabular}{c c c c}
\hline\hline 
Source ID & $z$ & ${\rm FWHM}_{\rm maj} \times {\rm FWHM}_{\rm min}$\tablefootmark{a} & P.A.\tablefootmark{b}\\[1ex]
          &     & [$\arcsec$] & [$\degr$] \\[1ex]
\hline 
AzTEC/C1a & 4.7\tablefootmark{c} & $0.40^{+0.04}_{-0.05} \times 0.31^{+0.04}_{-0.04}$ & $113.7^{+19.2}_{-21.1}$ \\ [1ex]
AzTEC/C2a & 3.179\tablefootmark{c} & $0.19^{+0.03}_{-0.04} \times 0.18^{+0.03}_{-0.04}$ & $166.4^{+35.5}_{-40.0}$ \\ [1ex]
AzTEC/C2b & $1.10^{+2.60}_{-1.10}$ & $0.32^{+0.04}_{-0.03} \times 0.16^{+0.03}_{-0.03}$ & $28.4^{+8.6}_{-7.5}$  \\ [1ex]
AzTEC/C3a\tablefootmark{d} & 1.125\tablefootmark{c} & $0.32^{+0.02}_{-0.02} \times 0.16^{+0.02}_{-0.02}$ & $132.9^{+5.7}_{-5.5}$ \\ [1ex]
AzTEC/C3c\tablefootmark{d} & $2.03^{+1.19}_{-0.31}$ & $0.29^{+0.05}_{-0.06} \times 0.12^{+0.05}_{-0.10}$ & $88.1^{+13.7}_{-17.0}$  \\ [1ex]
AzTEC/C4 & $5.30^{+0.70}_{-1.10}$ & $0.40^{+0.05}_{-0.05} \times 0.19^{+0.04}_{-0.04}$ & $4.0^{+8.7}_{-8.2}$ \\ [1ex]
AzTEC/C5 & 4.3415\tablefootmark{c} & $0.31^{+0.02}_{-0.03} \times 0.18^{+0.02}_{-0.03}$ & $99.4^{+7.1}_{-7.4}$  \\ [1ex]
AzTEC/C6a & 2.494\tablefootmark{c} & $0.20^{+0.03}_{-0.04} \times 0.15^{+0.03}_{-0.04}$ & $171.4^{+39.1}_{-30.8}$ \\ [1ex]
AzTEC/C6b & 2.513\tablefootmark{c} & $0.27^{+0.07}_{-0.08} \times 0.21^{+0.07}_{-0.08}$ & $65.9^{+44.4}_{-38.2}$  \\ [1ex]
AzTEC/C7 & $3.06^{+1.88}_{-1.76}$ & $0.35^{+0.03}_{-0.02} \times 0.10^{+0.02}_{-0.03}$ & $59.6^{+3.5}_{-3.7}$    \\ [1ex] 
AzTEC/C8a & 3.62\tablefootmark{c} & $0.21^{+0.07}_{-0.13} \times 0.17^{+0.11}_{-0.13}$ & $127.0^{+36.8}_{-42.9}$    \\ [1ex]
AzTEC/C9a & $2.68^{+0.24}_{-0.51}$ & $0.29^{+0.02}_{-0.03} \times 0.08^{+0.03}_{-0.06}$ & $157.0^{+4.7}_{-4.9}$  \\ [1ex] 
AzTEC/C9b & 2.8837\tablefootmark{c} & $0.37^{+0.06}_{-0.08} \times <0.09$ & $132.9^{+7.7}_{-7.3}$  \\ [1ex] 
AzTEC/C9c & 2.9219\tablefootmark{c} & $0.12^{+0.05}_{-0.05} \times <0.09$ & $85.6^{+24.0}_{-43.8}$ \\ [1ex]
AzTEC/C10a & $3.40^{+3.60}_{-0.59}$ & $0.50^{+0.08}_{-0.09} \times <0.09$ & $0.4^{+5.5}_{-5.3}$  \\ [1ex]
AzTEC/C10b & $2.90^{+0.30}_{-0.90}$ & $0.38^{+0.07}_{-0.07} \times 0.16^{+0.05}_{-0.06}$ & $73.6^{+10.4}_{-11.9}$  \\ [1ex]
AzTEC/C11\tablefootmark{d} & $4.30^{+0.07}_{-3.33}$ & $0.31^{+0.02}_{-0.03} \times 0.21^{+0.02}_{-0.03}$ & $95.5^{+10.9}_{-12.4}$ \\ [1ex]
AzTEC/C12 & $3.25^{+0.16}_{-0.51}$ & $0.45^{+0.05}_{-0.05} \times 0.14^{+0.04}_{-0.03}$ & $56.3^{+4.6}_{-4.7}$ \\ [1ex]
AzTEC/C13a & $2.01^{+0.15}_{-0.49}$ & $0.58^{+0.07}_{-0.07} \times 0.11^{+0.04}_{-0.07}$ & $6.0^{+3.7}_{-3.6}$  \\ [1ex]
AzTEC/C14 & $4.58^{+0.25}_{-0.68}$ & $0.62^{+0.05}_{-0.06} \times 0.11^{+0.03}_{-0.04}$ & $18.1^{+2.4}_{-2.3}$\\ [1ex]
AzTEC/C15 & $3.91^{+0.28}_{-2.35}$ & $0.24^{+0.03}_{-0.04} \times 0.20^{+0.04}_{-0.03}$ & $53.2^{+40.2}_{-44.9}$ \\ [1ex]
AzTEC/C16a & $3.15^{+0.62}_{-1.54}$ & $0.62^{+0.18}_{-0.19} \times 0.33^{+0.12}_{-0.13}$ & $110.4^{+28.3}_{-20.5}$ \\ [1ex]
AzTEC/C16b & $2.39^{+0.27}_{-0.56}$ & $0.22^{+0.04}_{-0.09} \times 0.19^{+0.08}_{-0.04}$ & $55.1^{+44.5}_{-42.8}$ \\ [1ex]
AzTEC/C17 & 4.542\tablefootmark{c} & $0.30^{+0.03}_{-0.04} \times 0.12^{+0.03}_{-0.05}$ & $153.1^{+7.2}_{-7.4}$ \\ [1ex]
AzTEC/C18 & $3.15^{+0.13}_{-0.44}$ & $0.43^{+0.05}_{-0.05} \times 0.27^{+0.04}_{-0.04}$ & $112.6^{+11.2}_{-11.4}$  \\ [1ex] 
AzTEC/C19 & $2.87^{+0.11}_{-0.41}$ &  $0.20^{+0.03}_{-0.02} \times 0.11^{+0.03}_{-0.02}$ & $51.2^{+12.9}_{-11.7}$   \\ [1ex]
AzTEC/C20 & $3.06^{+0.13}_{-0.54}$ & $0.17^{+0.03}_{-0.05} \times 0.13^{+0.03}_{-0.05}$ & $56.5^{+43.2}_{-41.7}$   \\ [1ex]
AzTEC/C21 & $2.70^{+1.30}_{-0.40}$  & $0.42^{+0.06}_{-0.06} \times 0.15^{+0.04}_{-0.06}$ & $89.9^{+6.7}_{-7.0}$\\ [1ex]
AzTEC/C22a & 1.599\tablefootmark{c} & $0.19^{+0.03}_{-0.02} \times 0.13^{+0.03}_{-0.03}$ & $127.8^{+18.4}_{-15.6}$\\ [1ex]
AzTEC/C22b & 1.599\tablefootmark{c} & $0.27^{+0.05}_{-0.06} \times <0.09$ & $87.4^{+11.3}_{-12.8}$ \\ [1ex]
AzTEC/C23 & $2.10^{+0.46}_{-0.41}$ & $0.70^{+0.18}_{-0.19} \times 0.20^{+0.09}_{-0.11}$ & $158.7^{+9.1}_{-9.0}$ \\ [1ex] 
AzTEC/C24a & $2.01^{+0.19}_{-0.46}$ & $0.27^{+0.05}_{-0.05} \times 0.13^{+0.05}_{-0.05}$ & $16.5^{+16.2}_{-13.7}$ \\ [1ex]
AzTEC/C24b\tablefootmark{d} & $2.10^{+0.08}_{-0.63}$ & $0.29^{+0.10}_{-0.13} \times 0.15^{+0.10}_{-0.15}$ & $151.1^{+33.2}_{-34.9}$ \\ [1ex]
AzTEC/C25 & 2.51\tablefootmark{c} & $0.45^{+0.11}_{-0.12} \times 0.13^{+0.07}_{-0.13}$ & $70.3^{+10.5}_{-11.0}$ \\ [1ex]
AzTEC/C26 & $5.06^{+0.08}_{-0.90}$ & $0.47^{+0.07}_{-0.07} \times 0.14^{+0.04}_{-0.07}$ & $123.4^{+6.2}_{-6.2}$\\ [1ex] 
AzTEC/C27 & $2.77^{+0.88}_{-0.47}$ & $0.31^{+0.10}_{-0.11} \times 0.20^{+0.08}_{-0.12}$ & $9.1^{+40.1}_{-17.9}$ \\ [1ex]
\hline
\end{tabular} 
\tablefoot{The sources AzTEC/C1b, C8b, C10c, and C13b were not detected in our ALMA 870~$\mu$m survey, while AzTEC/C3b was not covered by our ALMA pointings (see Sect.~2).\tablefoottext{a}{Deconvolved FWHM of the major and minor axes derived through Gaussian fits in the image plane using the AIPS task {\tt JMFIT}.}\tablefoottext{b}{Major axis position angle of the fitted Gaussian measured from north through east. Formally, the P.A. lies in the range ${\rm P.A.} \in [0\degr, \, 180\degr]$, but some of the tabulated values have uncertainties that place the P.A. value being outside this range. However, the P.A. is symmetrical under a $180\degr$ rotation.}\tablefoottext{c}{Spectroscopic redshift (see \cite{brisbin2017}, and references therein).}\tablefoottext{d}{No {\tt MAGPHYS} SED could be derived for the source, while AzTEC/C11 and C24b were excluded from the SED analysis because they are likely to host an active nucleus (M17b).}   }
\end{minipage} }
\end{table}

\begin{figure*}[!htb]
\begin{center}
\includegraphics[width=\textwidth]{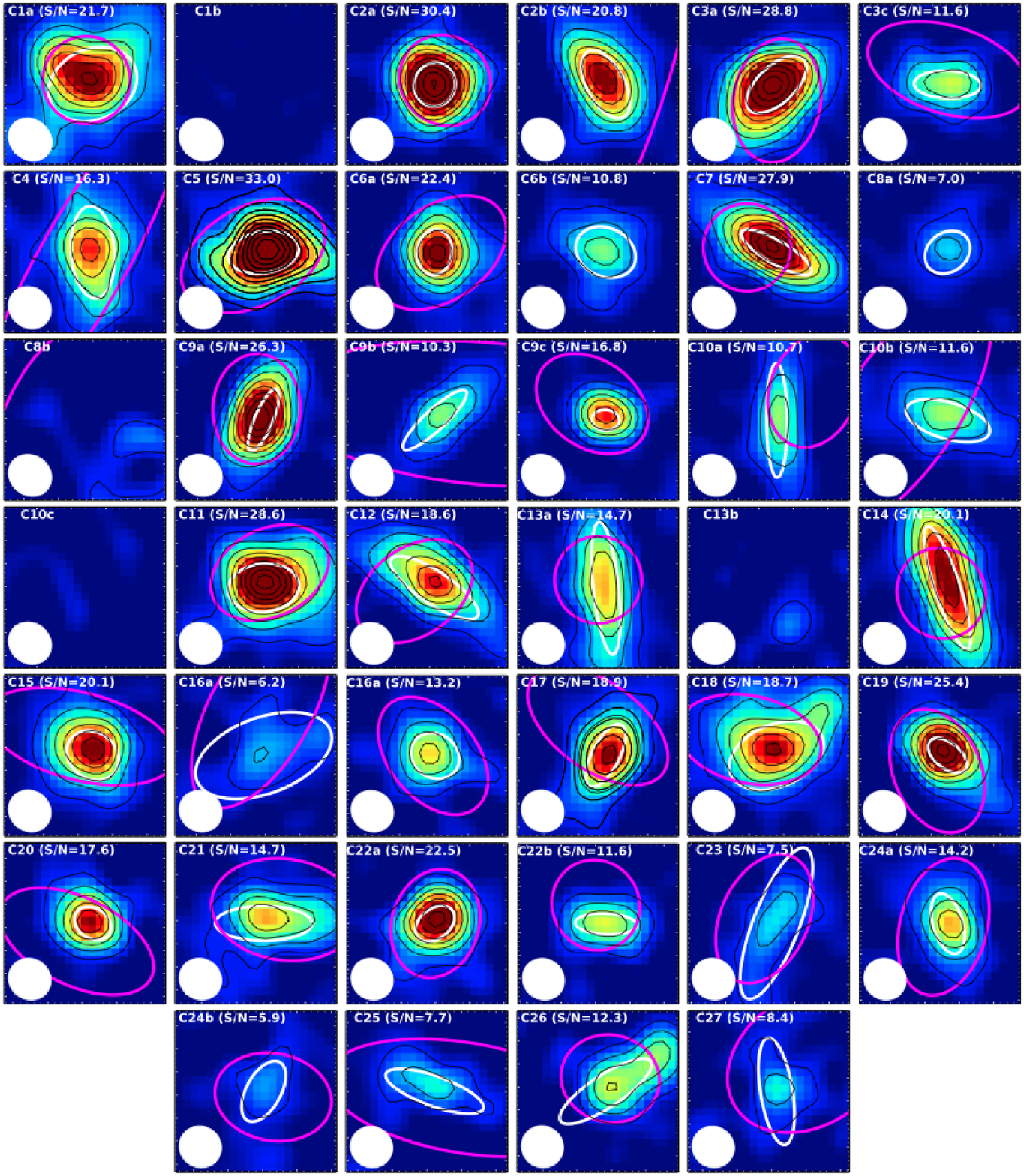}
\caption{Observed-frame 870~$\mu$m ALMA images towards AzTEC/C1--C27. Each image is centred on the ALMA 870~$\mu$m peak position 
(except the non-detections (AzTEC/C1b, C8b, C10c, and C13b), which are centred on the ALMA 1.3~mm position), is $0\farcs7 \times 0\farcs7$ in size, 
oriented such that north is up and east is left, and displayed in a common linear colour-scale. 
The contour levels start from $3\sigma$, and progress in steps of $3\sigma$. The detection ${\rm S/N}_{\rm 870\, \mu m}$ ratio is indicated 
in parenthesis. The white and magenta ellipses show the deconvolved FWHM source sizes at 870~$\mu$m and 3~GHz (the present study and M17a, 
respectively). The ALMA synthesised beam FWHM is shown in the bottom left of each panel.}
\label{figure:maps}
\end{center}
\end{figure*}

The ALMA 870~$\mu$m images towards AzTEC/C1--C27 are shown in Fig.~\ref{figure:maps}, and the derived source 
sizes are tabulated in Table~\ref{table:sample}. In Table~\ref{table:data}, we list the values of the 
binned average data points ($\Sigma_{\rm gas}$ and $\Sigma_{\rm SFR}$) plotted in Fig.~\ref{figure:ks}.

\begin{table}[H]
\renewcommand{\footnoterule}{}
\caption{Molecular gas and SFR surface densities of the binned average data points shown in Fig.~\ref{figure:ks}.}
{\normalsize
\begin{minipage}{1\columnwidth}
\centering
\label{table:data}
\begin{tabular}{c c}
\hline\hline 
$\log(\Sigma_{\rm gas}/{\rm M}_{\sun}\,{\rm pc}^{-2})$\tablefootmark{a} & $\log(\Sigma_{\rm SFR}/{\rm M}_{\sun}\,{\rm yr}^{-1}\,{\rm kpc}^{-2})$  \\[1ex]
\hline 
\multicolumn{2}{c}{MS objects\tablefootmark{b} }\\ [1ex]
\hline 
$3.89\pm0.14$ & $1.26\pm0.07$\\ [1ex]
$4.30\pm0.05$ & $1.60\pm0.13$\\ [1ex]
$4.63\pm0.04$ & $1.86\pm0.16$\\ [1ex]
$4.88\pm0.05$ & $2.03\pm0.34$\\ [1ex]
\hline 
\multicolumn{2}{c}{Starbursts\tablefootmark{c}}\\ [1ex]
\hline 
$4.14\pm0.06$ & $1.64\pm0.26$\\ [1ex]
$4.43\pm0.05$ & $2.03\pm0.20$\\ [1ex]
$4.84\pm0.05$ & $2.09\pm0.13$\\ [1ex]
$5.03\pm0.07$ & $2.64\pm0.20$\\ [1ex]
\hline
\end{tabular} 
\tablefoot{\tablefoottext{a}{The gas masses used to derive these gas surface densities were estimated using the Scoville et al. (2016) 
dust continuum method, which is based on the assumption of a uniform, Galactic $\alpha_{\rm CO}$ conversion factor.}\tablefoottext{b}{The MS definition was adopted from Speagle et al. (2014).}\tablefoottext{c}{The starbursts were defined as objects that lie above the MS mid-line by a factor of $>3$.}}
\end{minipage} }
\end{table}

\section{K-S diagrams constructed using different CO-to-H$_2$ conversion factors}

\begin{figure}[!htb]
\centering
\resizebox{0.98\hsize}{!}{\includegraphics{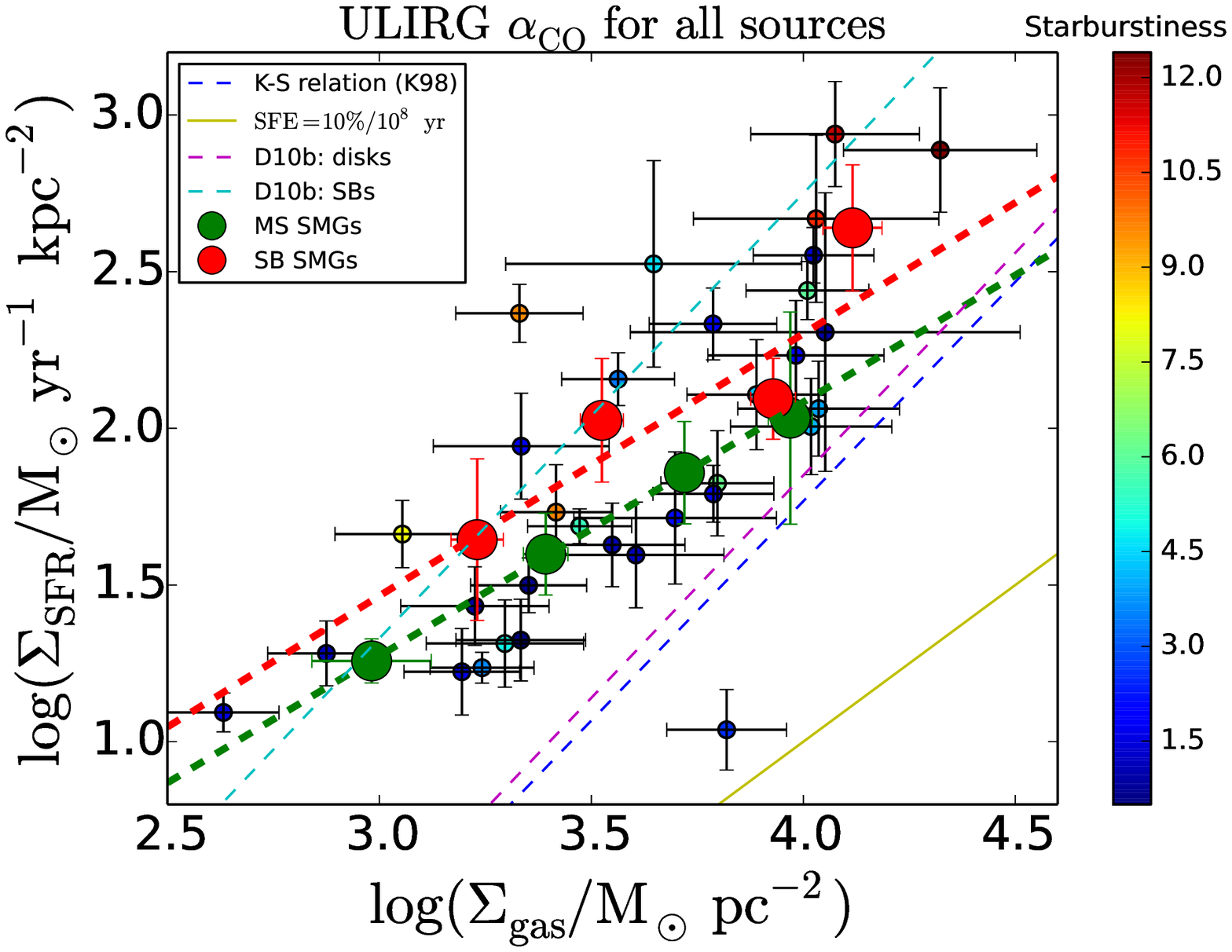}}
\resizebox{0.98\hsize}{!}{\includegraphics{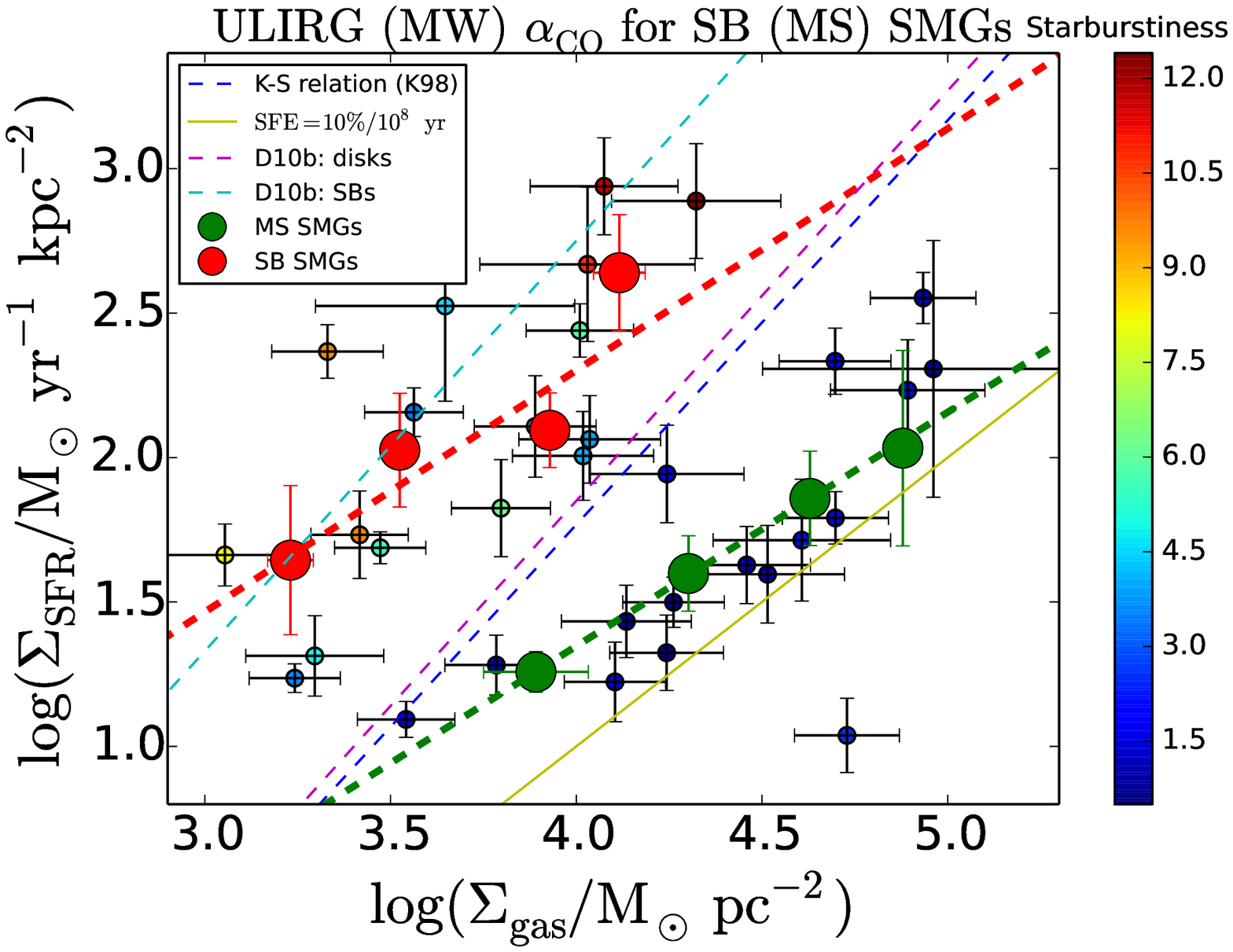}}
\caption{\textbf{Top:} Similar to Fig.~\ref{figure:ks}, but all the $\Sigma_{\rm gas}$ values were calculated by 
scaling the dust-based gas masses by a factor of $0.8/6.5$ to make them consistent with a ULIRG $\alpha_{\rm CO}$ conversion factor 
of 0.8~M$_{\sun}$~(K~km~s$^{-1}$~pc$^2$)$^{-1}$. \textbf{Bottom:} Similar to the top panel, but only the starburst SMGs' $\Sigma_{\rm gas}$ values 
were calculated by using the aforementioned ULIRG $\alpha_{\rm CO}$ factor, while a Galactic value was assumed for the MS objects. 
The plotting ranges of the two panels are different for legibility purposes.}
\label{figure:ksulirg}
\end{figure}

In the top panel in Fig.~\ref{figure:ksulirg}, we show a similar K-S diagram to that in Fig.~\ref{figure:ks}, but where all the $\Sigma_{\rm gas}$ values were calculated by assuming a ULIRG $\alpha_{\rm CO}$ factor of 0.8~M$_{\sun}$~(K~km~s$^{-1}$~pc$^2$)$^{-1}$. The linear least squares fits through the binned averages yielded the slope and $y$-intercept of ($a = 0.81\pm 0.01$, $b = -1.16 \pm 0.04$) for the MS SMGs, and ($a = 0.84 \pm 0.39$, $b = -1.04 \pm 1.49$) for the starburst SMGs. The slopes remain the same as in Fig.~\ref{figure:ks}, but the former (latter) normalisation is higher by a factor of 5.37 (5.89). This makes most of our average starburst data points consistent with the D10b starburst sequence.

The K-S diagram shown in the bottom panel in Fig.~\ref{figure:ksulirg} was constructed by assuming the same Galactic $\alpha_{\rm CO}$ factor for the MS SMGs as in Fig.~\ref{figure:ks}, and the aforementioned ULIRG-like factor for starbursts. This creates a clear bimodal distribution in the K-S plane (starbursts versus MS objects). The corresponding best-fit parameters for the MS SMGs are the same as in Fig.~\ref{figure:ks} ($a = 0.81\pm 0.01$, $b =  -1.89 \pm 0.05$), and for the starbursts they are the same as quoted above ($a = 0.84 \pm 0.39$, $b = -1.04 \pm 1.49$).

\section{K-S diagrams constructed using the 3~GHz sizes}

\begin{figure}[!htb]
\centering
\resizebox{0.98\hsize}{!}{\includegraphics{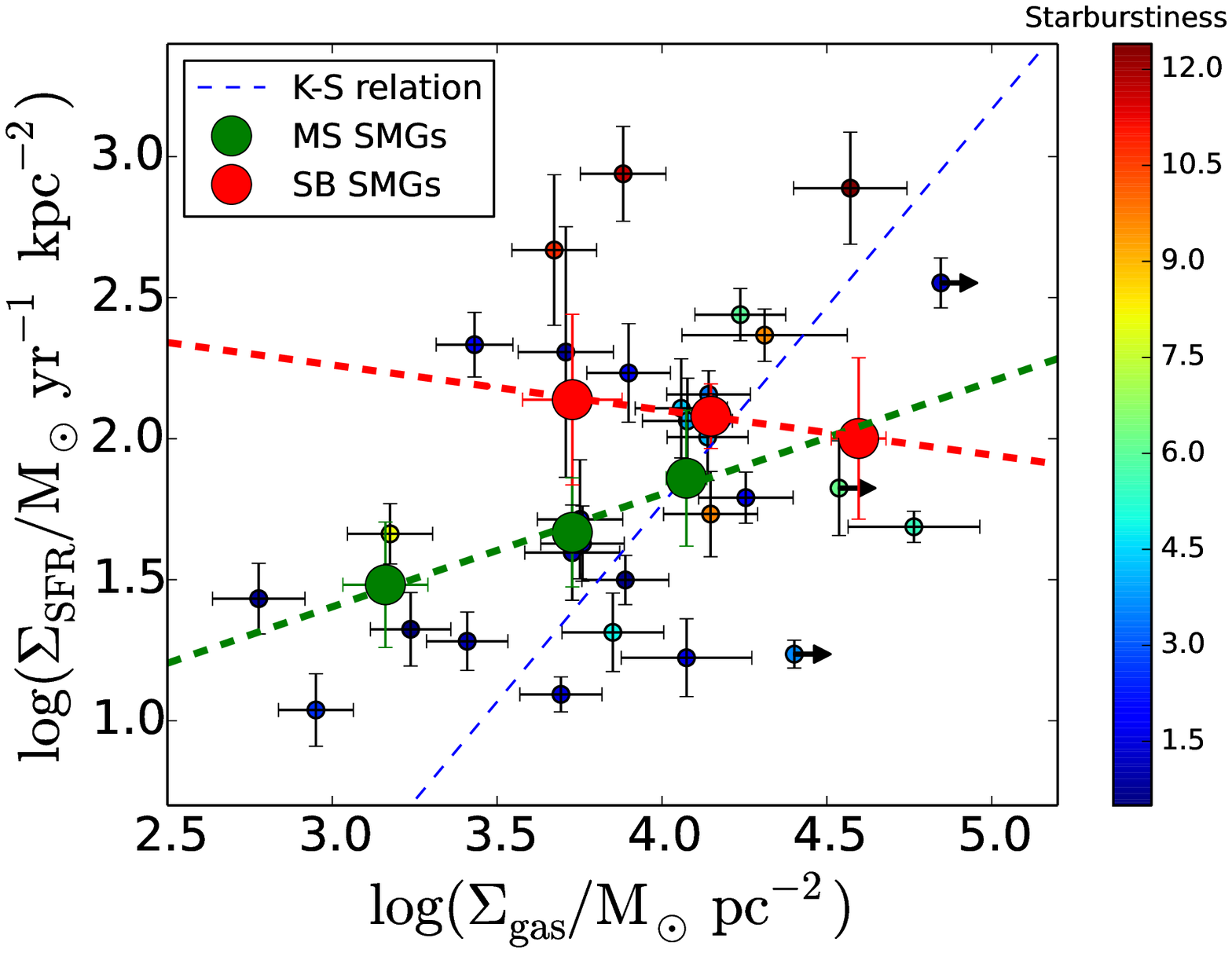}}
\resizebox{0.98\hsize}{!}{\includegraphics{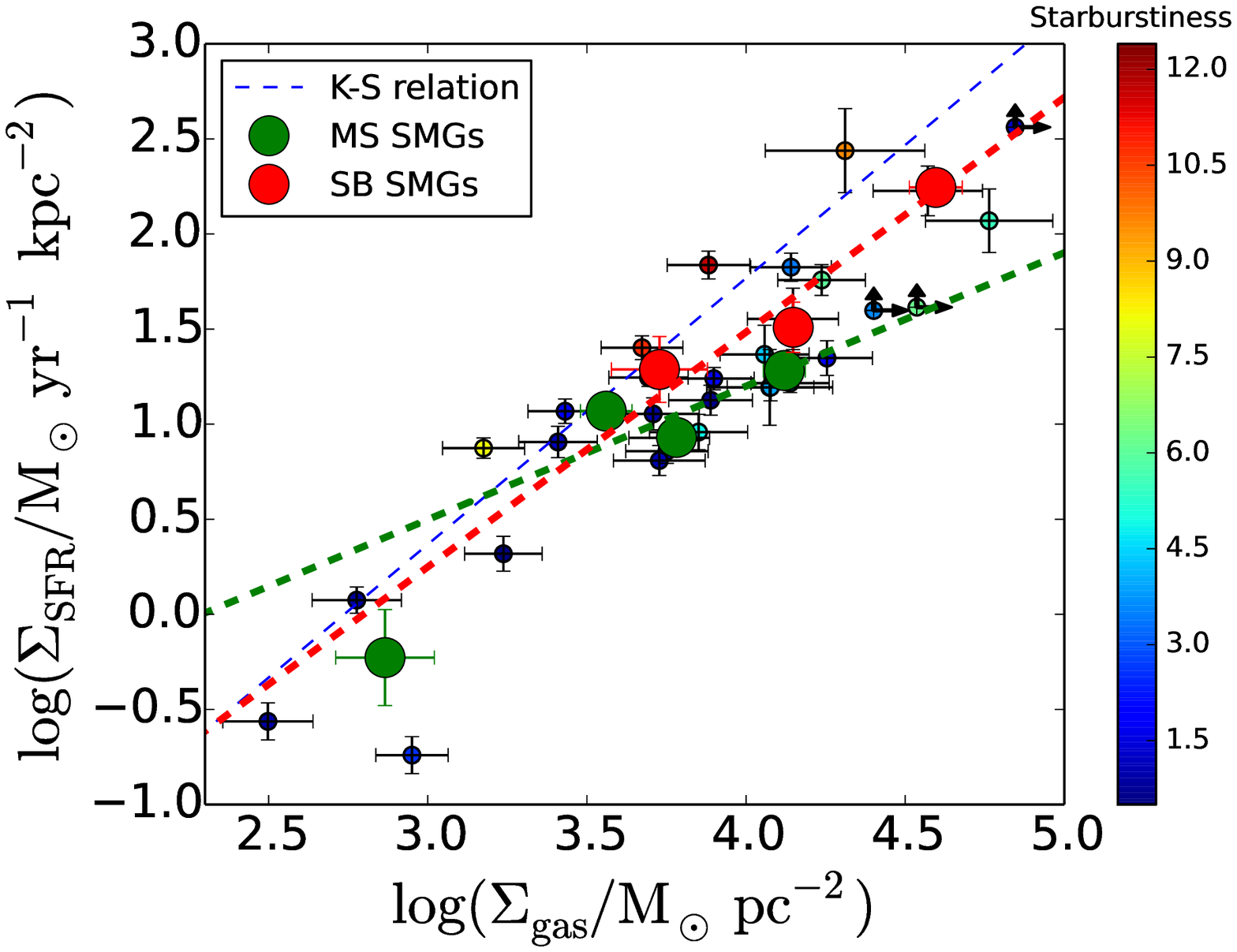}}
\caption{\textbf{Top:} Similar to Fig.~\ref{figure:ks}, but $\Sigma_{\rm gas}$ was calculated over the 3~GHz radio sizes derived by M17a (the magenta ellipses in Fig.~\ref{figure:maps}). Each bin contains five sources. \textbf{Bottom:} Similar to the top panel, but both $\Sigma_{\rm gas}$ and $\Sigma_{\rm SFR}$ were calculated over the 3~GHz radio sizes. Each MS (SB) bin contains four (five) sources, where the one additional source compared to the top panel is the 3~GHz detected SMG AzTEC/C8b. In both panels, the three sources unresolved at 3~GHz (lower limit to $\Sigma_{\rm gas}$ in the top panel, and to both $\Sigma_{\rm gas}$ and $\Sigma_{\rm SFR}$ in the bottom panel) were incorporated into the binned averages using a right-censored K-M survival analysis. The K98 relationship is shown for comparison. The plotting ranges of the two panels are different for legibility purposes.}
\label{figure:ks_radio}
\end{figure}

In the top panel in Fig.~\ref{figure:ks_radio}, we show a modified version of Fig.~\ref{figure:ks} where the gas surface densities were calculated over the 3~GHz radio-emitting sizes (M17a; see the magenta ellipses in Fig.~\ref{figure:maps}). The K-S diagram shown in the bottom panel in Fig.~\ref{figure:ks_radio} has both the SFR and gas surface densities calculated over the 3~GHz sizes. The data were binned separately for the MS and starburst objects, and the three sources that were unresolved at 3~GHz (AzTEC/C1a, C7, and C13a) were incorporated into the binned averages using a right-censored Kaplan-Meier (K-M) survival analysis (see M17a for details). The linear least squares fit parameters were found to be $(a=0.40\pm0.07,\,b=0.20\pm0.24)$ for the MS SMGs, and $(a=-0.16\pm0.02,\,b=2.74\pm0.07)$ for the starbursts in the top panel. The corresponding parameters for the data plotted in the bottom panel are $(a=0.70\pm0.30,\,b=-1.61\pm1.20)$ and $(a=1.23\pm0.29,\,b=-3.45\pm1.29)$, respectively. The results suggest that $\Sigma_{\rm SFR}$ and $\Sigma_{\rm gas}$ should be compared over common size scales (K98). However, as discussed in M17a, the 3~GHz radio emission might not always be probing the spatial extent of active high-mass star formation (and hence $\Sigma_{\rm SFR}$), but instead the radio-emitting region can be puffed up as a result of the same galaxy interaction that triggers the SMG phase. Hence, in the main text we focused on the K-S relation derived using the 870~$\mu$m dust-emitting sizes.

\end{document}